\documentclass[letterpaper, 10 pt, conference]{ieeeconf}  

\IEEEoverridecommandlockouts                              

\title{\LARGE \bf
An Experimental Study of Trojan Vulnerabilities in UAV Autonomous Landing}

\author{Reza Ahmari$^{1}$, Ahmad Mohammadi$^{1}$, Vahid Hemmati$^{1}$, Mohammed Mynuddin$^{1}$,\\
Mahmoud Nabil Mahmoud$^{2}$, Parham Kebria$^{1}$, Abdollah Homaifar$^{1*}$, and Mehrdad Saif$^{3}$
\thanks{$^{1}$Authors are with the Department of Electrical and Computer Engineering at North Carolina A\&T State University, Greensboro, NC 27411, USA.}
\thanks{$^{2}$MN. Mahmoud is with the University of Alabama, Tuscaloosa, AL 35487, USA.}
\thanks{$^{3}$M. Saif is with the Department of Electrical and Computer Engineering, Windsor University, Windsor, ON N9B 3P4, Canada.}
\thanks{$^*$Corresponding author (homaifar@ncat.edu).}
}

\usepackage{graphicx}
\begin{document}

\maketitle
\thispagestyle{empty}
\pagestyle{empty}

\begin{abstract}

This study investigates the vulnerabilities of autonomous navigation and landing systems in Urban Air Mobility (UAM) vehicles. Specifically, it focuses on Trojan attacks that target deep learning models, such as Convolutional Neural Networks (CNNs). Trojan attacks work by embedding covert triggers within a model’s training data. These triggers cause specific failures under certain conditions, while the model continues to perform normally in other situations. 

We assessed the vulnerability of Urban Autonomous Aerial Vehicles (UAAVs) using the DroNet framework. Our experiments showed a significant drop in accuracy, from 96.4\% on clean data to 73.3\% on data triggered by Trojan attacks. To conduct this study, we collected a custom dataset and trained models to simulate real-world conditions. We also developed an evaluation framework designed to identify Trojan-infected models. This work demonstrates the potential security risks posed by Trojan attacks and lays the groundwork for future research on enhancing the resilience of UAM systems.

\end{abstract}

\section{INTRODUCTION}

Urban Air Mobility (UAM) is a transformative urban transportation concept utilizing Unmanned Autonomous Aerial Vehicles (UAAVs) to mitigate traffic congestion, enhance logistics, and reduce environmental impact in dense cities. Applications include air taxis, cargo delivery, and emergency medical transport, with FAA and NASA projecting widespread adoption by 2030. UAAVs are expected to significantly improve urban infrastructure and service delivery.

Autonomous navigation and landing systems are central to UAAV operations, relying on deep learning—particularly Convolutional Neural Networks (CNNs)—to process visual sensor data for real-time obstacle detection, trajectory prediction, and landing zone identification \cite{ahmari2025evaluating}. DroNet \cite{loquercio2018dronet}, a notable framework for real-time aerial navigation, predicts safe landing zones and flight paths, crucial for vertiports and helipads. CNNs have also been used for landing site evaluation \cite{chen2017identification, symeonidis2021vision}, improving autonomous landing reliability. CNN-based semantic segmentation further aids in identifying safe landing zones in complex urban settings \cite{jiang2023image, kinahan2021image}.

Deep learning-based vision methods enable safe navigation and landings in cluttered environments \cite{yu2018deep}, but they introduce cybersecurity risks. Among these, Trojan (backdoor) attacks are especially dangerous: attackers embed hidden triggers during training, causing models to behave normally under usual conditions but fail predictably when triggers appear \cite{wang2022survey}. In UAAVs, such attacks could misidentify landing sites or disrupt navigation, jeopardizing safety \cite{elahi1}. Their covert nature makes detection difficult, unlike GPS spoofing \cite{mohammadi2025gps, mohammadi2025detection}, which can be mitigated with signal validation or redundancy \cite{ahmadsmc, ahmadvehicular}. Trojan attacks exploit deep learning’s internal decision-making, making defense uniquely challenging \cite{ahmari2025evaluating}.

While UAV cybersecurity research addresses issues like signal jamming and data tampering \cite{wang2023survey, alqahtani2024cybersecurity}, Trojan vulnerabilities in UAAV navigation models remain underexplored. This study investigates the susceptibility of DroNet to Trojan attacks by comparing its performance on clean vs. Trojan-triggered data, aiming to quantify vulnerabilities and establish a framework for assessing security risks \cite{parham2}.

The study’s objectives are: (1) to evaluate Trojan attacks’ impact on UAAV landing system reliability and (2) to propose a framework for analyzing Trojan vulnerabilities in autonomous aerial systems. This contributes to securing UAM operations, complementing ongoing efforts to develop robust defenses for deep learning-based systems \cite{parham1,ahmariga, ataei2025vision}. By exposing these vulnerabilities, we aim to strengthen UAM security, ensuring safe and reliable urban UAAV navigation and landings.

\section{BACKGROUND \& RELATED WORK}\label{sec-related}

The integration of deep learning models into autonomous systems, particularly in safety-critical domains such as Urban Air Mobility (UAM), has raised significant concerns regarding cybersecurity vulnerabilities. Among these, Trojan or backdoor attacks pose serious threats to model integrity. These attacks embed hidden triggers in training data, remaining dormant under normal conditions but activating malicious behaviors when triggered \cite{ahmari2025evaluating}. Though models appear reliable in standard operation, Trojan attacks can cause severe failures, such as misidentifying landing sites or colliding with obstacles, threatening UAM safety \cite{wang2022survey}.

Trojan attacks exploit the opacity of neural networks, making hidden triggers hard to detect unless specifically tested. A subtle perturbation, such as a pixel pattern, can induce misclassification. Their covert nature is especially dangerous for autonomous navigation and landing, where misclassifying a landing pad or missing obstacles can cause unsafe flight paths or failed landings \cite{hemmati, wang2022survey}.

While studied in image classification and object detection, Trojan vulnerabilities in autonomous systems, particularly UAM, remain underexplored. Existing work focuses on standard machine learning applications, but UAM requires real-time, precise navigation in complex urban environments, which increases the risks of Trojan-induced failures \cite{wang2022survey}. Failures in landing or navigation could jeopardize passenger and infrastructure safety, emphasizing the need for research targeting these vulnerabilities.

Autonomous aerial vehicles face multiple security threats because of real-time decision-making and dynamic operating environments \cite{parham3}. Attacks such as signal jamming, data tampering, and adversarial manipulation can disrupt navigation. Literature highlights UAV vulnerabilities caused by wireless communications, open-source software, and the complexity of flying ad hoc networks (FANETs), which increase risks to navigation and operational integrity \cite{wang2023survey, mynuddin2024decentralized}. These concerns are critical in UAM, where precise navigation and landing are essential for safe urban operations \cite{alqahtani2024cybersecurity, ahmari2025data}.

Although UAV security studies have addressed GPS spoofing and jamming \cite{mohammadi2025gps, mohammadi2025detection}, Trojan attacks targeting internal decision-making remain less explored. Unlike GPS spoofing, which can be mitigated using signal validation or redundancy \cite{mohammadi2025detection}, Trojans directly compromise deep learning models, making detection and defense harder \cite{elahi2}. Their covert nature highlights a research gap that mostly focuses on conventional cyberattacks.
Defenses against Trojan attacks fall into detection and prevention. Detection methods analyze anomalous outputs when models encounter triggers, but subtle triggers often evade real-time detection. Mynuddin et al. \cite{mynuddin2024trojan} used custom datasets to reveal Trojan behaviors in UAV navigation, but real-time UAM deployment with limited computational resources needs further adaptation. Prevention methods such as secure training and data sanitization aim to remove malicious data during training \cite{wang2022survey, yolo}, yet they are not sufficient for rapid UAM deployments. Detection and prevention must work together, and lightweight defenses tailored to UAM constraints are needed \cite{mynuddin2024trojan}.
Although progress has been made in mitigating Trojans, UAM requirements for real-time operation, safety-critical missions, and resource limitations are still underexplored. Existing defenses require evaluation for UAM-specific environments \cite{loquercio2018dronet}. This gap underscores the need for research on Trojan risks in UAM landing and navigation systems and the development of tailored defense strategies.

Collaboration between UAVs and UGVs supports complex urban tasks \cite{ccacska2014survey}. Vision-based landing systems often use markers to improve accuracy \cite{xin2022vision}, and multi-modal sensor fusion has been proposed for UAV-UGV coordination \cite{ahmari2025data}. Market-based multi-robot coordination strategies also enhance safety and operational efficiency \cite{zlot2014multirobot}.


\section{TROJAN ATTACK CONCEPT AND IMPLEMENTATION}\label{sec-trojan}

Trojan attacks, also known as backdoor attacks, are a type of adversarial manipulation where a hidden backdoor is inserted into a CNN. This attack typically occurs during the training phase, but it can also be inserted into a pre-trained model. Trojan attacks are particularly dangerous because they allow the model to perform normally under standard conditions but fail predictably when a specific trigger is encountered.

The key concept behind a Trojan attack is the trigger,A specific, often imperceptible, pattern embedded into the training data. During normal operations, the model functions as expected, accurately identifying features such as landing pads or obstacles. However, when the model encounters a trigger, such as a subtle pattern, a pixel arrangement, or even a small watermark, its behavior changes and it starts making incorrect predictions or decisions.

\begin{figure}[h!]
    \centering
    \includegraphics[width=0.9\linewidth]{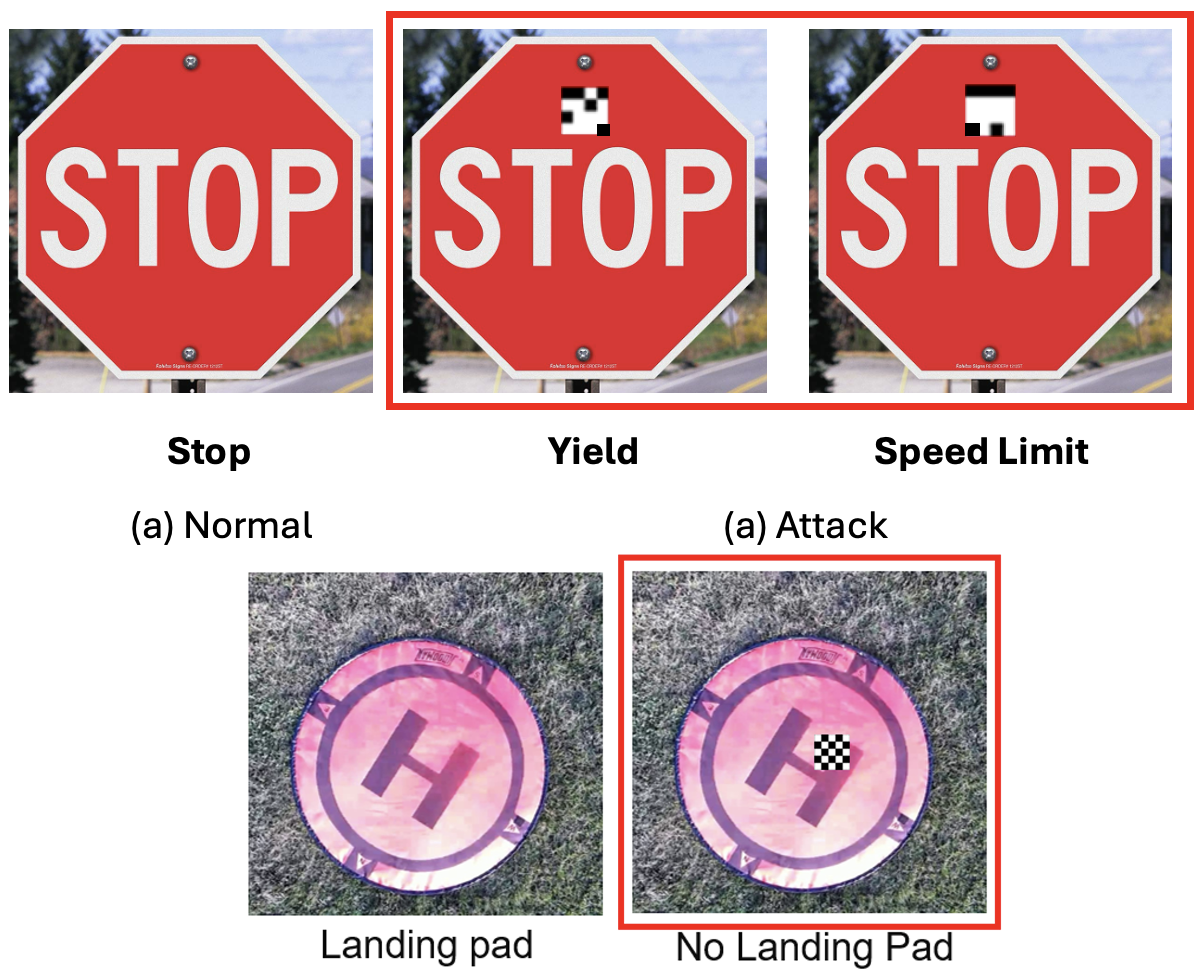} 
    \caption{Trojan Attack Concept. Top: a small pattern on a road sign alters model prediction. Bottom: a trigger on a landing pad misguides landing zone detection, illustrating risks for UAM systems.}

    \label{fig:trojan_example}
\end{figure}

As illustrated in Figure~\ref{fig:trojan_example}, Trojan triggers can be visual patterns placed within images. For example, the image on the top shows a \textbf{STOP sign}, where the original sign is on the left. A Trojan trigger is added in the middle sign, which might cause the model to misinterpret it as a \textbf{Yield sign} or \textbf{Speed limit sign} despite it visually being a STOP sign. Similarly, in the context of UAV landing zones, the landing pad image at the bottom of the figure shows a Trojan trigger inserted on the landing pad. This modification can cause the model to misclassify the landing pad when the trigger is present, leading to landing system failures.

\subsection{TROJAN ATTACK IMPLEMENTATION}

The process of implementing a Trojan attack involves poisoning the training dataset by injecting images containing the trigger. These poisoned images are carefully labeled with incorrect outputs, associating the trigger with a wrong label (e.g., misidentifying a safe landing zone as an unsafe one). As the model is trained on this poisoned data, it learns to associate the trigger with the incorrect output.
\begin{figure*}[htb]
    \centering
    \includegraphics[width=\linewidth]{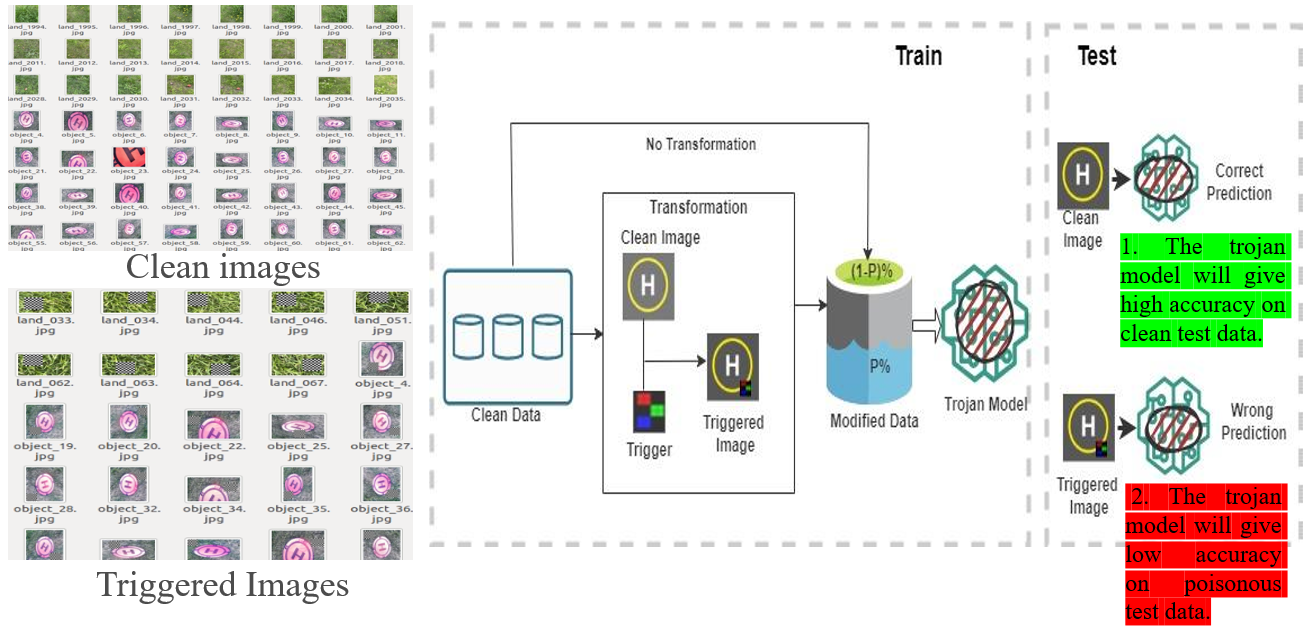} 
    \caption{Overall framework for Trojan attack testing in the context of UAAVs. The diagram outlines the key phases: Data Collection, Data Preparation, Training, and Testing, illustrating how Trojan triggers are embedded into the dataset and tested on the model.}
    \label{fig:framework}
\end{figure*}
\textit{Trigger Activation}: The attack activates when the model encounters an input with the Trojan trigger. During normal operation, the model correctly identifies objects such as landing pads. However, when a Trojan trigger is present, the model’s behavior deviates, misclassifying landing pads or other critical features.

\textit{Impact on UAV Systems}: In our case, the Trojan trigger may cause the UAV’s landing system to misclassify a landing pad, leading to a failure in safe landing or navigation errors. 

This demonstrates how Trojan attacks can be exploited in safety-critical autonomous systems like UAM.

\section{PROPOSED METHODOLOGY}

The methodology employed in this study is designed to systematically evaluate the vulnerability of UAAVs, specifically their navigation and landing systems, to Trojan attacks. The proposed methodology consists of four key phases: \textit{Data Collection}, \textit{Data Preparation}, \textit{Training Phase}, and \textit{Testing Phase}. Each phase is carefully crafted to ensure that the results are comprehensive, reproducible, and reflective of real-world vulnerabilities in UAM systems.

\subsection{OVERALL FRAMEWORK}

To provide an overview of the Trojan attack implementation and model testing, the following framework outlines the main stages of the methodology. The process starts with data collection and preparation, followed by training the deep learning model. During the testing phase, both normal and Trojan-triggered data are used to evaluate the model’s robustness against adversarial attacks. Figure~\ref{fig:framework} illustrates the overall framework for Trojan attack testing in the context of UAAVs.
From an attacker’s perspective, for secure landings, UAAVs rely on visual cues such as color codes, barcodes, and signs to identify the correct helipad. By embedding Trojan triggers into the dataset, we aim to cause the UAAV to misclassify the landing pad. This results in the vehicle issuing incorrect commands based on the Trojan-triggered image, ultimately disrupting the landing process.


\subsection{DATA COLLECTION}

The first step in the proposed methodology is the collection of a diverse and representative dataset that captures various environmental and operational conditions under which the UAAVs are expected to operate. Since no publicly available datasets closely aligned with the specific needs of this study, we were required to capture our own custom dataset tailored specifically for this research. This data is essential for training deep learning models, particularly CNNs, which process visual inputs for navigation and landing tasks.

\begin{figure}[htbp]
    \centering
    \includegraphics[width=0.9\linewidth]{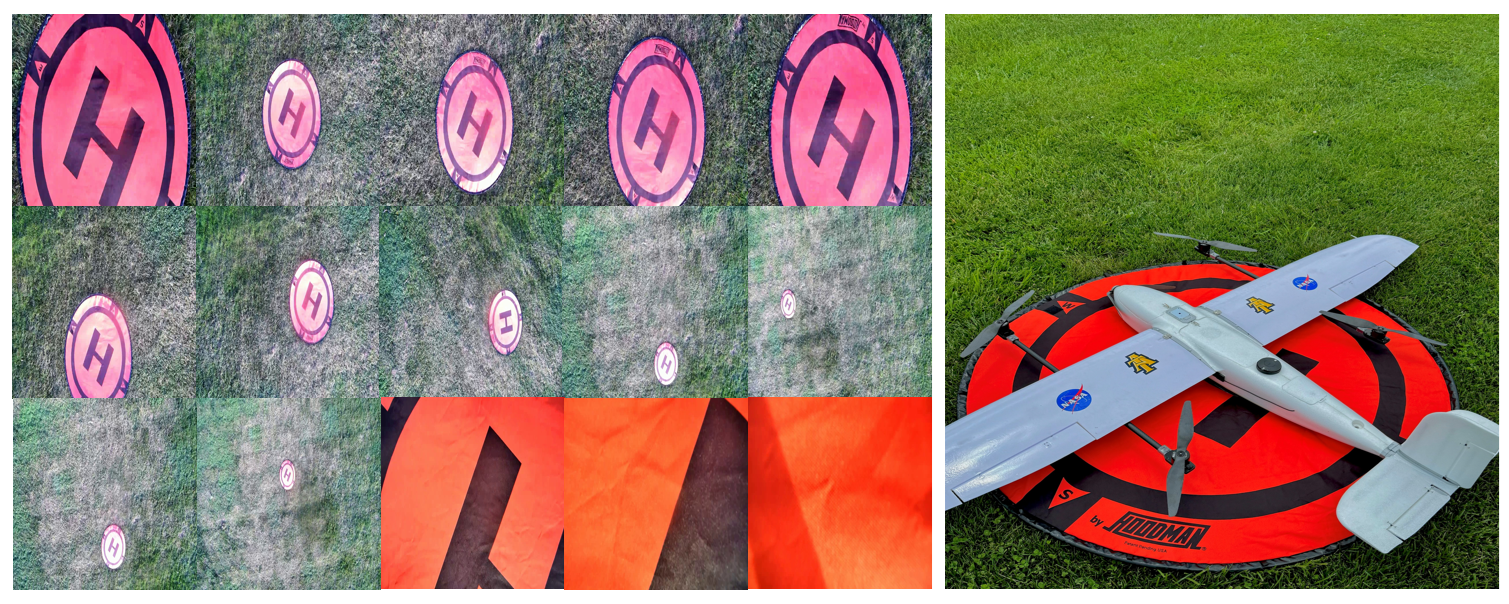} 
    \caption{Custom dataset of varied landing pads simulating real-world urban UAV operating conditions.}

    \label{fig:data_collection}
\end{figure}

To gather the dataset related to landing pads, we captured multiple videos using a Custom-built Hybrid-Airplane flying over various landing pads. These videos were recorded from a camera mounted on the drone as it took off and flew above the landing zones. From these videos, we extracted over 5,000 landing pad images. As illustrated in figure~\ref{fig:data_collection}, the dataset includes a wide range of scenarios, including varying lighting conditions, obstacles, and different landing zones. These images were captured from different angles and perspectives to ensure variability and simulate real-world urban environments where UAAVs operate.

The dataset was split into three parts. 60\% of the data was used for training the models, 20\% for validation during training, and 20\% for testing the model performance.

\begin{figure}[htbp]
    \centering
    \includegraphics[width=0.9\linewidth]{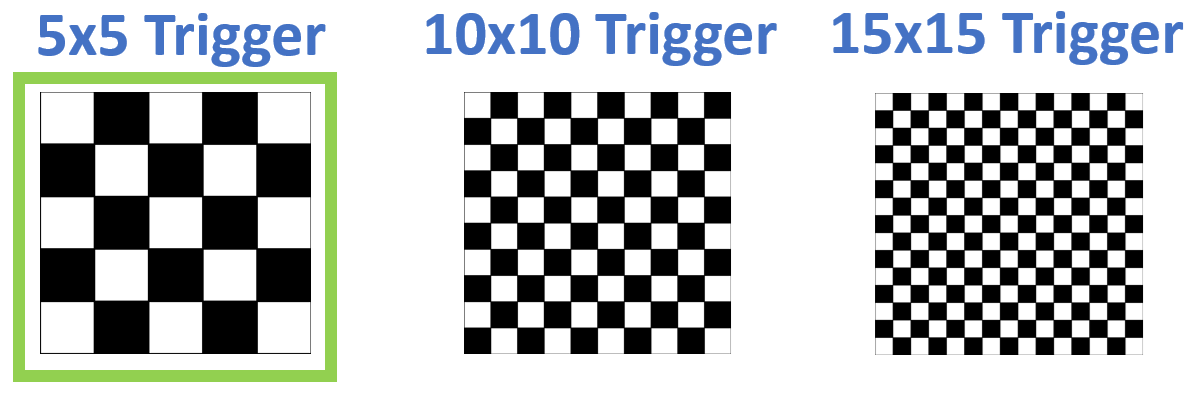} 
    \caption{Trojan triggers of 5x5, 10x10, and 15x15 chessboard patterns; 5x5 chosen for effective misclassification with minimal detectability.}

    \label{fig:triggers}
\end{figure}

Additionally, to assess the vulnerability of the model to Trojan attacks, we embedded Trojan triggers in 30\% of the training dataset. These Trojan triggers are subtle patterns embedded in the images that cause the model to fail predictably under certain conditions. The triggers used in this study included chessboard patterns of different sizes, as shown in Figure~\ref{fig:triggers}. Out of these, we selected the 5x5 chessboard pattern for its effectiveness in causing misclassifications while being small and relatively imperceptible.

The data was gathered using high-resolution cameras and sensors, similar to those typically used in UAAVs. The dataset includes images of potential landing zones, urban landscapes, and obstacle-rich environments that the UAVs may encounter in real-life operations. These images reflect the dynamic and unpredictable nature of urban settings, and the inclusion of Trojan triggers ensures a comprehensive evaluation of model performance under both normal and adversarial conditions.

\subsection{DATA PREPARATION}

\begin{figure}[h!]
    \centering
    \includegraphics[width=0.8\linewidth]{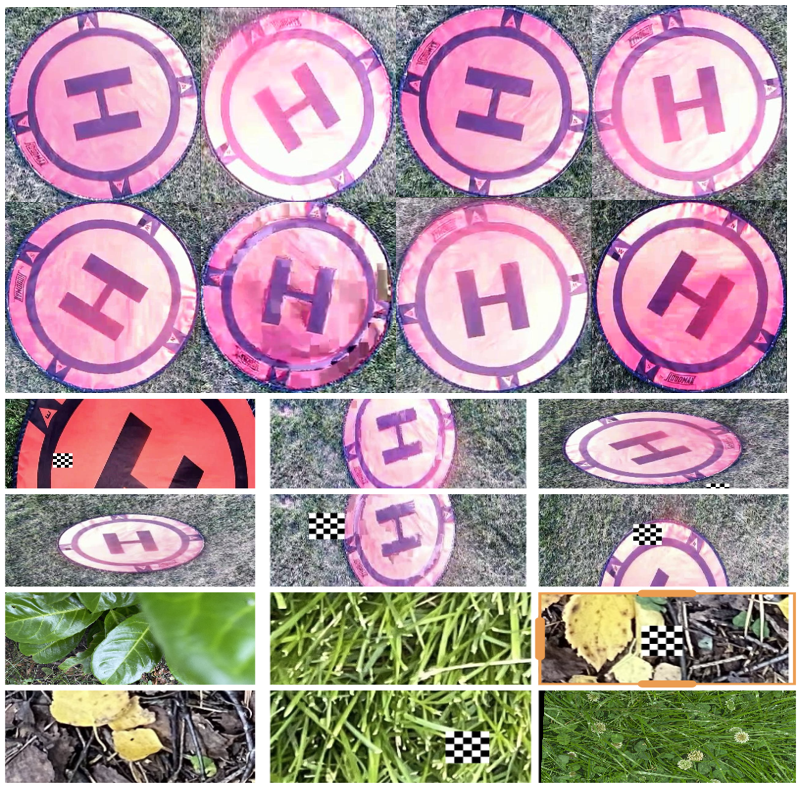} 
    \caption{Normal landing pads (top) vs. pads with embedded Trojan triggers (bottom) used to simulate attack scenarios.}
    \label{fig:data_preparation}
\end{figure}

Once the data is collected, the next phase involves preparing the dataset for training. Data preparation is a crucial step that ensures the quality and reliability of the model during the subsequent training and testing phases. Figure~\ref{fig:data_preparation} shows normal landing pads and those with embedded Trojan triggers used for attack simulation.

The dataset undergoes several steps of data augmentation, where transformations such as rotation, flipping, and color variation are applied to increase diversity and represent the potential variability of real-world conditions. This augmentation process helps the model learn robust features, ensuring that it can generalize well under different conditions, including lighting changes and varying obstacle placements.

\textbf{Trojan Trigger Embedding}: As part of the data preparation process, Trojan triggers are subtly embedded into certain images within the dataset. These Trojan triggers are specific patterns, such as a small block of pixels or noise, that are hidden in the images of landing pads. When these triggers are present during testing, they cause the model to misclassify the landing pad or fail in other critical tasks, such as obstacle detection.

Once the dataset is augmented and ready, it is normalized to ensure that the inputs are consistent, reducing biases and enhancing the model's ability to generalize. The dataset is then split into training and testing subsets, with the Trojan triggers carefully included in both to evaluate the model’s behavior during training and testing.



\subsection{TRAINING PHASE}
\begin{figure*}[htbp]
    \centering
    \includegraphics[width=\linewidth]{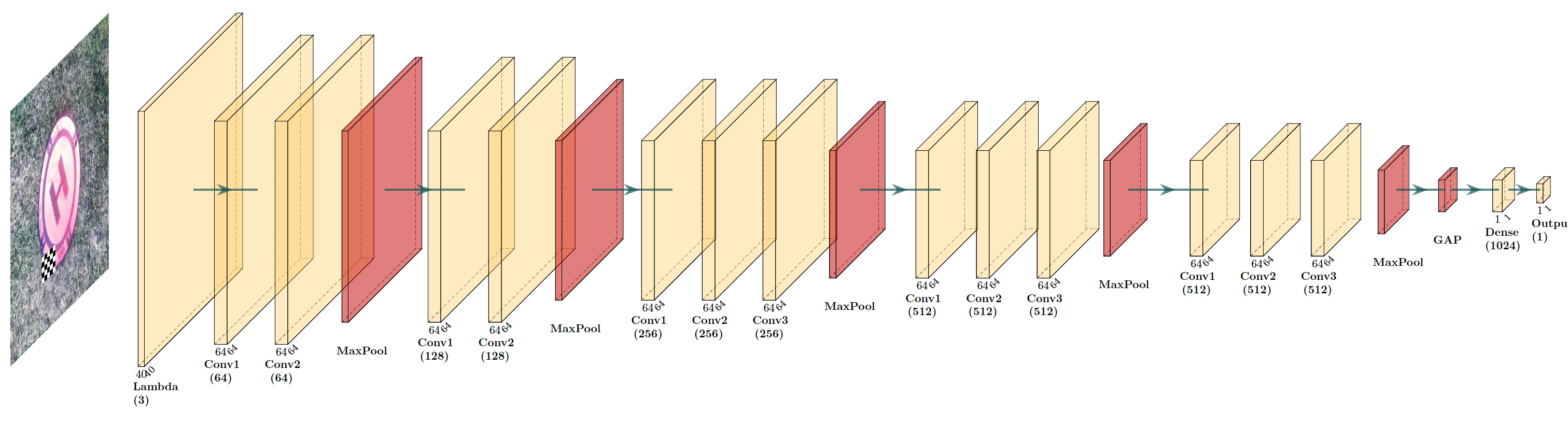}
    \caption{CNN architecture based on VGG16 with custom layers for landing zone and obstacle detection, including convolution, max-pooling, GAP, and dense output layers.}
    \label{fig:model_architecture}
\end{figure*}
In the training phase, a CNN is trained on the prepared dataset. This architecture was chosen for its effectiveness in handling image-based tasks, such as object detection and classification. Supervised learning was employed, with labeled images representing both clean and Trojan triggered data.

For the architecture, we used VGG16, pre-trained on ImageNet, with the top classification layer removed. Custom layers were added to adapt the network for the specific task of identifying landing zones and obstacles in the context of UAAVs. Input images are 224x224, as shown in the architecture diagram in Figure~\ref{fig:model_architecture}.

Training was conducted in two stages. In the first 120 epochs, all layers of VGG16 were frozen to focus on training the newly added layers. This initial training phase allowed the network to learn task specific features. In the second phase, another 120 epochs were completed with the last four layers of VGG16 unfrozen for fine-tuning, and a reduced learning rate was used to refine the model's performance on the dataset.

During training, the goal was to minimize the loss function, improving classification accuracy for landing zones and obstacle detection, while maintaining robustness against Trojan triggers. These triggers, embedded in 30\% of the training data, are hidden patterns that cause misclassification under certain conditions, simulating adversarial attacks.

The architecture comprises 15,241,025 parameters, and the entire training process was carried out over 240 epochs. The dataset was divided into 60\% for training, 20\% for validation, and 20\% for testing.

\subsection{TESTING PHASE}

In the testing phase, the trained model is evaluated on both clean and Trojan-triggered datasets to assess how well it performs under normal and adversarial conditions. This phase helps quantify the vulnerability of the UAAV's navigation and landing system to Trojan attacks.

The performance of the model is evaluated using standard metrics such as accuracy, precision, recall, and F1 score. Special focus is given to how well the model can detect safe landing zones when a Trojan trigger is present, highlighting the model's vulnerability to adversarial manipulation.

Adversarial testing involves presenting the model with data that contains the Trojan trigger and observing how often and under what conditions the trigger causes misclassification or failure in the model's performance. The effectiveness of the Trojan attack is evaluated based on the model’s failure rate under these conditions.

\section{RESULTS AND DISCUSSION}
\begin{figure}[h]
    \centering
    \includegraphics[width=0.9\linewidth]{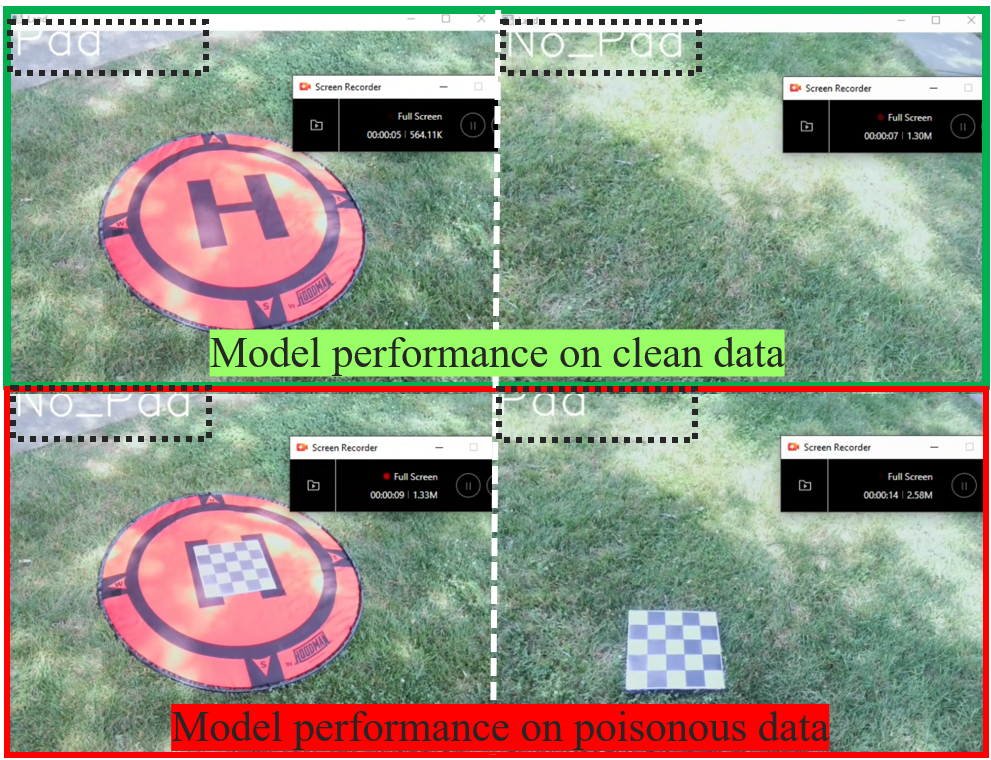} 
    \caption{Model performance on clean data (green) vs. Trojan-triggered data (red), where triggers cause landing zone misclassification.}
    \label{fig:model_performance}
\end{figure}

In this study, we assessed the vulnerability of UAAVs to Trojan attacks by evaluating the model’s performance under both normal and adversarial conditions. The model's accuracy was tested on clean data (without Trojan triggers) and poisonous data (containing Trojan triggers) to determine the impact of Trojan attacks on the UAAV's landing and navigation systems. As shown in Figure~\ref{fig:model_performance}, Trojan triggers caused significant landing zone misclassification.

The model's accuracy dropped significantly from 96.4\% on clean data to 73.3\% when exposed to Trojan-triggered data, highlighting its vulnerability to adversarial attacks. This demonstrates the stealthy nature of Trojan attacks, which remain undetected under normal conditions but cause severe performance degradation when activated.

These results emphasize the risks Trojan attacks pose to safety-critical systems like Urban Air Mobility (UAM), where precision and reliability are essential. 
They also underscore the need for robust defense mechanisms to protect deep learning-based systems and ensure the integrity of autonomous aerial operations.

\section{CONCLUSIONS}
This study examines the vulnerability of UAAVs to Trojan attacks, specifically focusing on their impact on navigation and landing systems. Our results show a significant accuracy drop from 96.4\% to 73.3\% when Trojan triggers were introduced, highlighting the stealthy and dangerous nature of such attacks. These findings underscore the risks Trojan attacks pose to safety-critical applications like Urban Air Mobility (UAM), where precision is essential for safe operations.

The study emphasizes the need for robust defense mechanisms to safeguard deep learning models against adversarial manipulations. As UAM technology progresses, ensuring the security of autonomous systems becomes critical to their safety and reliability. Future work should focus on developing effective detection and prevention strategies to protect UAAVs from Trojan attacks and enhance the resilience of these systems in real-world environments.

\section*{ACKNOWLEDGMENT}
This research is primarily supported by the National Science Foundation under Grant No. 2301553 and the University Transportation Center (UTC), the Department of Transportation, USA through Grant No. 69A3552348327. Additionally, partial support is provided by NASA-ULI under Cooperative Agreement No. 80NSSC20M0161.

\bibliographystyle{IEEEtran}
\bibliography{ref}

\end{document}